# Magnetic Dirac Fermions and Chern Insulator Supported on Pristine Silicon Surface

Huixia Fu[1], Zheng Liu[2], Chao Lian[1], Jin Zhang[1], Hui Li[1], Jia-Tao Sun[1]*, Sheng Meng[1,3]*

[1]Beijing National Laboratory for Condensed Matter Physics and Institute of Physics, Chinese Academy of Sciences, Beijing 100190, P. R. China

[2]Institute for Advanced Study, Tsinghua University, Beijing 100084, P. R. China

[3]Collaborative Innovation Center of Quantum Matter, Beijing 100190, P. R. China

## Abstract

Emergence of ferromagnetism in non-magnetic semiconductors is strongly desirable, especially in topological materials thanks to the possibility to achieve quantum anomalous Hall effect. Based on first-principles calculations, we propose that for Si thin film grown on metal substrate, the pristine Si(111)-$\sqrt{3} \times \sqrt{3}$ surface with a spontaneous weak reconstruction has a strong tendency of ferromagnetism and nontrivial topological properties, characterized by spin polarized Dirac-fermion surface states. In contrast to conventional routes relying on introduction of alien charge carriers or specially patterned substrates, the spontaneous magnetic order and spin-orbit coupling on the pristine silicon surface together gives rise to quantized anomalous Hall effect with a finite Chern number $C = -1$. This work suggests exciting opportunities in silicon-based spintronics and quantum computing free from alien dopants or proximity effects.

*Address correspondence to smeng@iphy.ac.cn; jtsun@iphy.ac.cn

## I. INTRODUCTION

Being able to generate magnetism in semiconducting materials that are not naturally magnetic brings new opportunities to spintronics applications. Recently the pursuit of ferromagnetic ordering in the two dimensional (2D) $d^0$ materials has attracted wide attention owing to the intriguing physical phenomena and tremendous potential applications. In graphene and BN, besides directly doping transition metal elements, the magnetic moments can also be induced by introducing nonmagnetic adatoms [1], vacancy defects, structural distortions [2-4], edge engineering [5-7], and the proximity with ferromagnetic substrates [8]. However, these dopants and defects often induce rather localized magnetic moments; robust long-range ordered ferromagnetism can rarely be obtained in theoretical proposals and experimental measurements [9-14].

Silicon is the most popular $d^0$ element for semiconductors. The intrinsic ferromagnetism in silicon-based materials could hold great promise for new magnetoelectric effects and nanoscale spintronics, naturally compatible with current silicon industry. A prototypical approach to achieve magnetism is to decorate alien atoms on Si(111) [15-20]. Numerous works have demonstrated the complex spin patterns in Si(111) covered with 1/3 monolayer Sn/Pb atoms, which was attributed to strong electronic correlation induced by adatoms [15-19]. Moreover, Erwin *et al*. reported long-range spin polarization at graphitic steps on a family of vicinal silicon surfaces [20]. Despite that ferromagnetism is believed to originate intrinsically from silicon surfaces, these stepped surfaces have to be stabilized by a small amount of extrinsic gold atoms. Novel properties of Si (111) surfaces in addition to magnetism have also been widely explored [21-23]. Wang *et al*. proposed that the 1/3 monolayer halogen adatoms on Si(111) can create anisotropic Dirac bands [21]. A *p*-band-element *X* decorated Au/Si(111) surfaces were reported as a promising prototype for larger-gap topological states [22]. The quantum anomalous Hall (QAH) states were also predicted in the W/Cl-covered Si(111) systems [23]. However, in all these proposals, alien elements are always necessary to give rise to magnetic and topological orders mentioned above, thus bringing additional difficulties in controlled doping and surface stability. To date, long range ordered magnetic states in pristine silicon systems have not been reported either in experiment or theory. Not to mention achieving topological electronic structures simultaneously.

In this letter, we propose that ferromagnetism can emerge on weakly reconstructed pure Si(111)-$\sqrt{3}\times\sqrt{3}$R30° surface (hereafter designated as Si-$\sqrt{3}$ surface) by first-principles calculations. This state breaks time-reversal symmetry and is characterized by the spin-polarized Dirac subbands with partially filled $p_z$ orbitals, originated from the interplay between spontaneous surface reconstruction and magnetic instability. More remarkably, we found that a band gap of 15 meV is opened up by spin-orbit coupling (SOC) interactions, resulting in a nontrivial QAH state characterized by a Chern number of $\mathcal{C}=-1$ and chiral edge states. In contrast to conventional magnetic topological insulators such as doped dilute magnetic semiconductors and tetradymite semiconductors [24], the spontaneous reconstructed Si-$\sqrt{3}$ surface with spin-polarized nontrivial topological bands may serve as a new pathway to realize QAH effect [8, 23-25], without introduction of foreign magnetic impurities nor substrates. The proposed Si-$\sqrt{3}$ surface was arguably fabricated/can be prepared by epitaxial growth of Si thin film on metal substrate under low temperature condition in a controllable way to remove other competing phases, therefore the surface ferromagnetism could be experimentally accessible. However we note that the previous reports about "multilayer silicene" on Ag substrate with a comparable surface structures have not been unambiguously explained by the model studied here; adatom structures with silver (or other impurities) may account for the observed periodicity [26]. Moreover, the low stability may make the surface reactive to unintended surface contamination, which may easily destroy the fragile electronic structure.

## II. COMPUTATIONAL METHODS

The first-principles calculations were performed using Perdew-Burke-Ernzerhof (PBE) functional [27] in Vienna Ab-initio Simulation Package (VASP) [28]. Projector-augmented wave (PAW) pseudopotentials are applied with an energy cutoff at 400 eV. The Brillouin zone is sampled with an $11 \times 11 \times 1$ Monkhorst-Pack $k$-mesh. The Si(111) slabs of three to seven silicon layers with a surface lattice constant of 6.67 Å were chosen with a vacuum region $\geq 15$ Å. The silicon atoms on the bottom layer are saturated by hydrogen atoms. The atomic geometry was optimized until the force on each atom is less than 0.01 eV/Å. Beyond PBE functional, the screened exchange hybrid density functional by Heyd-Scuseria-Ernzerhof (HSE06) [29,30] was also adopted to confirm the electronic structures. To describe the spin-polarized Dirac-type surface state and make Chern number calculation computationally efficient, an effective tight-binding (TB) Hamiltonian was constructed by projecting Bloch states onto maximally localized Wannier functions (MLWFs) using VASP2WANNIER90 interface [31,32]. Then a dense $k$ mesh of $600 \times 600 \times 1$ over the entire Brillouin zone was employed to ensure the convergence of integral.

## III. RESULTS AND DISCUSSION

The atomic structure of Si(111) slab with in-plane $\sqrt{3} \times \sqrt{3}$ R30° supercell is first optimized by performing non-spin-polarized calculations as shown in Fig. 1(a, b). We found that weak reconstruction of rhombic Si-$\sqrt{3}$ surface without bond breaking spontaneously takes place on all Si(111) slabs with $\geq 3$ silicon layers. The 1/3 of the three-coordinated Si$_A$ (green) atoms on the Si-$\sqrt{3}$ surface are 1.2 Å higher than the four-coordinated Si$_C$ (orange) atoms. The remaining 2/3 three-coordinated Si$_B$ (red) atoms with one dangling bond are nearly flat lying at a height of ~0.24 Å above Si$_C$ atoms. The Si$_B$ atoms form a super-hexagonal honeycomb lattice with a $\sqrt{3} \times \sqrt{3}$ R30° periodicity with respect to the ideal Si(111)-1×1 lattice. Compared with the $sp^3$ bonding features of ideal bulk Si(111), the low-buckled Si-$\sqrt{3}$ surface consisting of only two Si$_B$ atoms within the $\sqrt{3} \times \sqrt{3}$-R30° superlattice resembles closely perfect planar graphene structure. One expects that the unsaturated $p_z$ orbital from each Si$_B$ atom might form $sp^2$ hybridization giving rise to a Dirac cone in the band gap region of bulk states.

Representative electronic band structure of the Si-$\sqrt{3}$ surface in the non-magnetic (NM)

phase is shown in Fig. 2(a). Indeed, a Dirac-like state (orange lines) around the Fermi level can be easily distinguished from the bulk states (blue shadow areas). The corresponding density of states (DOS), as well as the orbital projected DOS (Fig. 2(b)) exhibits a partially filled peak with a band width of 0.2 eV around the Fermi level, mainly originating from the $p_z$ orbitals of $Si_B$ surface atoms. This fact indicates the Dirac-like state has the electronic origin of delocalized π bonds attributed to the planar super-hexagonal bond topology of $Si_B$ atoms. Furthermore, this partially filled Dirac state with a narrow band width, which is characterized as a sharp van Hove singularity in DOS, suggests a strong interaction among the electrons lying at the Fermi level. It is likely to invoke Stoner magnetic instabilities and result in a magnetic ground state accompanied by spin-dependent Dirac states around the Fermi level.

Naturally, we then consider the potential long-range magnetic orders in the surface states of the Si-$\sqrt{3}$ surface. Both the ferromagnetic (FM) and antiferromagnetic (AFM) ordering have been considered (Fig. S1). Interestingly, we found that the Si-$\sqrt{3}$ surface favors a FM spin ordering, with the calculated magnetic moment of $1\,\mu_B$ per $\sqrt{3}$ unit cell. This implies that a pair of $Si_B$ atoms per $\sqrt{3}$ unit cell share a single π-bonding electron while the $Si_A$ atom is fully saturated with a lone pair of two electrons, thanks to the symmetry breaking between $Si_A$ and $Si_B$ atoms in the $\sqrt{3}$ unit cell . In contrast, the initial AFM spin configuration will spontaneously relax into the NM phase. The favorable FM configuration is further verified in the larger $2\sqrt{3} \times 2\sqrt{3}$ R30° and $3 \times 3$ supercells to eliminate the potential errors induced by small cell sizes. Meanwhile, the priority of FM phase over both NM and AFM cases is found for all Si(111) slabs of various thickness as long as the surface reconstruction occurs. Our spin-polarized calculations predict that the FM structure is 91 meV more stable than the NM phase per $\sqrt{3}$ supercell. The qualitative results have also been reproduced by the HSE06 hybrid functional, which is more accurate to describe the *s-p* hybridization. HSE06 functional presents an even larger energy difference of 204 meV per $\sqrt{3}$ cell between the FM and NM phases. Coincident with the foregoing assumption, the calculated band structure and DOS for the FM structure exhibits two spin-polarized Dirac cones with an energy splitting of 0.39 eV at K point as shown in Fig. 2(c,d).

It is noteworthy that spin-majority channel gives rise to a zero-gap semiconductor, whereas the spin-minority channel displays an indirect semiconducting band gap of 0.44 eV, demonstrating a unique spin semiconductor feature, as illustrated in Fig. S2. Since the Fermi velocity $v_F$ is a key parameter to describe the fundamental characters of Dirac materials, the magnified spin-majority Dirac cone at the energy window from -0.3 eV to 0.4 eV obtained from both PBE and HSE06 functional is plotted in Fig. S3. We see the sharp feature of the Dirac band is very similar using different functionals. The Fermi velocity $v_F$ is $0.47 \times 10^6$ m/s by fitting the linear dispersion branch of the valence band from HSE06 function, which is comparable to the value of $10^6$ m/s for electrons in suspended graphene [33]. The spin density distribution displayed in Fig. 2(e, f) demonstrates that the ferromagnetic feature can be mostly attributed to the $p_z$ orbital of super-hexagonal $Si_B$ pattern, which has the largest contribution for the spin-polarized Dirac states ( Fig. 2(d)).

The favorable FM phase over NM state on Si-$\sqrt{3}$ surface signifies the magnetic instability predicted above, which can be understood from the Stoner criterion in the non-interacting regime [34]. One can determine whether $N(E_F)U > 1$ applies, where $N(E_F)$ is the DOS at the Fermi level in the non-spin-polarized state, and $U$ is a measure of the strength of the exchange interaction between the two Dirac cone in the spin-polarized state. Substituting with the values from density functional theory (DFT) calculations, we obtain $N(E_F)U = 6.53/(eV \cdot unit\ cell) \times 0.39\ eV = 2.5 > 1$. In this simple Stoner model, the condition for the emergence of ferromagnetic order is fulfilled, suggesting a spin polarized state might take place on the Si-$\sqrt{3}$ surface.

Being a quick predictive tool, Stoner theory is in fact not always successful due to the absence of many body interactions [35]. In order to understand the underlying spin exchange, one needs also to evaluate the electron-electron interaction not captured within the non-interacting regime. Via transforming the Bloch wave function of the spin-polarized Dirac subbands into real space Wannier basis, we obtain the single orbital hopping in the mean field framework $H_0 = \sum_{i,j} t_{ij} c_i^\dagger c_j$, where $i, j$ denote the positions of different $Si_B$ atoms, and the $t_{ij}$ denotes the electron hopping between them. The obtained nearest neighbor (NN) hopping amplitude $t_1$ is -52.6 meV in Wannier basis. In the representation of the same Wannier basis, the Hamiltonian for the dominant electron-electron interaction can be expressed as [36],

$$H_{int} = \widetilde{U}_0 \sum_i n_{i\downarrow} n_{i\uparrow} + \widetilde{U}_1 \sum_{\langle ij \rangle} n_i n_j + \widetilde{J}_{ex} \sum_{\langle ij \rangle \alpha} c_{i,\alpha}^\dagger c_{j,-\alpha}^\dagger c_{i,-\alpha} c_{j,\alpha} \quad (1)$$

where $\langle ij \rangle$ and $\alpha$ denote the nearest neighbor (NN) pairs and spin index, respectively. $\widetilde{U}_0$, $\widetilde{U}_1$, $\widetilde{J}_{ex}$ are respectively on-site Hubbard repulsion, NN direct repulsion, and NN direct exchange (see supporting information). Sampling the Wannier basis using Monte Carlo method, we obtain $\widetilde{U}_0 = 6.404$ eV, $\widetilde{U}_1 = 2.461$ eV, $J_{ex} = 4.5$ meV. The condition $t_1 \ll \widetilde{U}_0$ will push the Si-$\sqrt{3}$ surface toward a strongly localized spin sate. The positive $J_{ex} = 4.5$ meV suggests a ferromagnetic phase with a curie temperature up to 52 K.

With a staggered magnetic flux emerging from the spin-polarized Dirac fermion and the intrinsic SOC, the Si-$\sqrt{3}$ surface may exhibit QAH effect featuring the nontrivial Chern numbers. To illustrate this feature, we interpolate the spin-polarized Dirac bands based on the atom-centered $p_z$-like MLWFs with the SOC effect turned on. The Wannier-interpolated energy bands are in excellent agreement with the DFT results around the Fermi level (Fig. 3(a)). Furthermore a small band gap of ~4.5 meV (with PBE functional) is opened at the touching point of the spin-majority Dirac cone. The SOC gap value is higher than that of monolayer silicene (1.55 meV), which is the only pristine-silicon-based topological insulator that has been proposed [37]. When the Fermi level lies inside this energy gap, the insulating state could be probably a Chern insulator, which can be characterized by the first Chern number $\mathcal{C}$ calculated by,

$$\mathcal{C} = \frac{1}{2\pi} \sum_n \int_{BZ} d^2 k \, \Omega_n \quad (2),$$

where $\Omega_n$ is the momentum-space Berry curvature for the $n$th band [38-40] as

$$\Omega_n(k) = -\sum_{n' \neq n} \frac{2\text{Im}\langle \psi_{nk} | v_x | \psi_{n'k} \rangle \langle \psi_{n'k} | v_y | \psi_{nk} \rangle}{(\varepsilon_{n'} - \varepsilon_n)^2} \quad (3).$$

The summation is over all occupied valence bands in the first Brillouin zone, and $v_{x(y)}$ is the velocity operator along the *x(y)* direction. The absolute value of $\mathcal{C}$ corresponds to the number of gapless chiral edge states along the edges of the Si-$\sqrt{3}$ surface.

The Berry curvature $\Omega_n(k)$ along the high symmetry direction Γ-K-M-K'-Γ shows two sharp spikes of the same sign located right at the K and K' points (Fig. 3(b)). The even function of $\Omega_n(k)$ for all the occupied states can be clearly seen in the two dimensional distribution in the entire Brillouin zone, which also shows vanishing $\Omega_n(k)$ away from both valleys (Fig. 3(c)). By integrating the Berry curvature $\Omega_n(k)$ in the entire Brillouin zone, we obtain the Chern number $\mathcal{C} = -1$ indicating one nontrivial edge state. As expected from the nonzero Chern number, the anomalous Hall conductivity shows a quantized charge Hall plateau of $\sigma_{xy} = \mathcal{C} e^2/h$ when the Fermi level is located in the insulating gap of the spin-majority Dirac cone (Fig. 3(d)). Via HSE06 functional, the width of the Hall plateau will increase to 15 meV, which is readily accessible in experimental conditions.

The existence of topologically protected chiral edge states is one of the most important signatures of the QAH effect (Fig. 4(a)). To further reveal the nontrivial topological nature of the Si-$\sqrt{3}$ surface, we calculated the nanoribbon structure having a width of 200 zigzag silicon chains ignoring the effects of edge reconstruction using a tight binding model built on the Wannier basis described above. As shown in Fig. 4(b), we clearly see that the nontrivial edge states (red line) connecting the valence and conduction bands cross the insulating gap of the spin-majority Dirac cone. The appearance of only one chiral edge state is consistent with the calculated Chern number $\mathcal{C} = -1$, confirming the nontrivial topological nature of the Si-$\sqrt{3}$ surface. The spin polarized Dirac Fermion mediated topological characters suggest the pure silicon surface proposed here is an intrinsic magnetic topological insulator, holding the potential to be a successor for generating QAH effect.

Recently, the Si(111) thin films with $\sqrt{3}$ surface reconstruction have been successfully fabricated on silver substrates via molecular beam epitaxy (MBE) growth in a ultrahigh vacuum chamber [41,42]. Although the exact surface composition is yet under debate [26], the advances in thin film synthesis offer a promising route for experimental realization of QAH effect based on Si-$\sqrt{3}$ surface. We also confirm that the magnetism of the Si-$\sqrt{3}$ surface survives upon thin film growth on Ag(111) substrate (Fig. 4(c)). The calculated magnetic moment of 0.48 $\mu_B$ per Si$_B$ atom for the Si-$\sqrt{3}$ surface on sliver substrate is slightly less than the value (0.50 $\mu_B$/Si$_B$) for the freestanding silicon films, which should be the result of a small amount of charge transfer from Si film to Ag substrate. The spin-polarized Dirac surface states can be easily distinguished from projected band structures displayed in Fig. 4(d). The touching points of two spin-polarized Dirac cones are located at the $\Gamma$ point due to the band folding, with an obvious energy splitting of 0.38 eV. The spontaneous intrinsic magnetization survives when the number of silicon layers in the slab is $\geq 3$, where the reconstructed surface is satisfactorily far away from the Ag substrate to eliminate mutual interactions. The direct detection of spin polarization in such a system is called on, which can be achieved by surface sensitive instruments such as spin-polarized scanning tunneling microscope, magneto-optic Kerr effect measurement, or superconducting quantum interference devices. Moreover, the surface ferromagnetism discussed here provides a notable electronic feature to distinguish pure Si-$\sqrt{3}$ surface from the nonmagnetic Si-Ag alloy phase, suggesting a new way to justify the debated surface composition of multilayer silicene on silver substrate.

Since the Si-$\sqrt{3}$ surface has a weak reconstruction, additional steps might be taken to protect the surface from impurity contamination. Thus, we propose to use monolayer BN with $\sqrt{7} \times \sqrt{7}$ periodicity as a chemically inert layer on the top of the $\sqrt{3}$ structure, which has only a very small lattice mismatch < 1%, shown in Fig. S4(a). The van der Waals (vdW) functional in the vdW-optB88 [43] scheme is used to optimize the vdW packed heterostructure. The minimum vertical distance between BN and Si films is found to be 3.2 Å, indicating a weak interaction between them. As shown in Fig. S4(b), the projected band structures on BN film are completely isolated from the Dirac bands of Si surface, indicating spin-polarized Dirac surface states survive under the protection of BN films.

## IV. CONCLUSION

In conclusion, we propose that the pristine Si-$\sqrt{3}$ surface is pushed to a magnetic ground state upon spontaneous weak reconstruction. The ferromagnetic state is characterized by the spin polarized Dirac fermions with a high Fermi velocity of $0.47 \times 10^6$ m/s. More interestingly, the QAH effect can be experimentally accessed on pristine silicon surface thanks to a nontrivial SOC gap of 15 meV. These results suggest the pristine Si-$\sqrt{3}$ surface to be a potential intrinsic magnetic topological insulator, calling for experimental proofs. We have found the novel electronic characteristics of spin-polarized Dirac Fermions can persist on 1/3 layer H-terminated Si(111) and similar Ge-$\sqrt{3}$ surfaces, implying that the magnetism discussed here is a general phenomenon and would stimulate future design of spintronic and magnetoelectronic devices based on $d^0$ elements. This work would bring a new opportunity to understand magnetism in $d^0$ element silicon-based materials and might give rise to broad applications in silicon-compatible spintronic and magnetoelectronic devices.

**Acknowledgements**

The authors thank Dr. Zijing Ding for helpful discussions. This work was supported by the National Basic Research Program of China (Grant Nos. 2013CBA01600, 2015CB921001, 2012CB921403), the Natural Science Foundation of China (Grant No. 61306114) and "Strategic Priority Research Program (B)" of the Chinese Academy of Sciences (Grant No. XDB07030100), Chinese Youth 1000 Talents Program.

**Figures:**

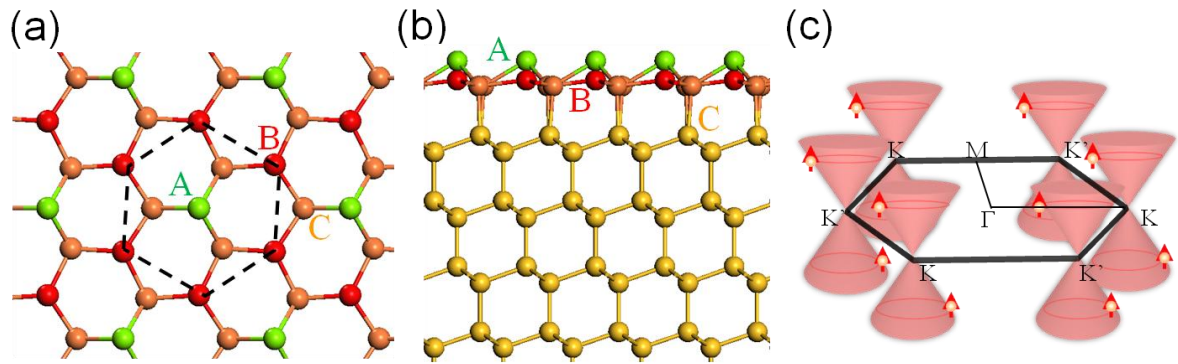

**Figure 1.** Top view (a) and side view (b) of Si-$\sqrt{3}$ surface. The green, red and orange balls respectively denote three types of silicon atoms: Si$_A$ (highest), Si$_B$ (lower), Si$_C$ (lowest) on the surface layer. Yellow spheres are the bulk Si atoms. The bottom Si layer is saturated by H atoms. (c) Schematic diagram of spin polarized Dirac cones around Fermi level in the first Brillouin zone.

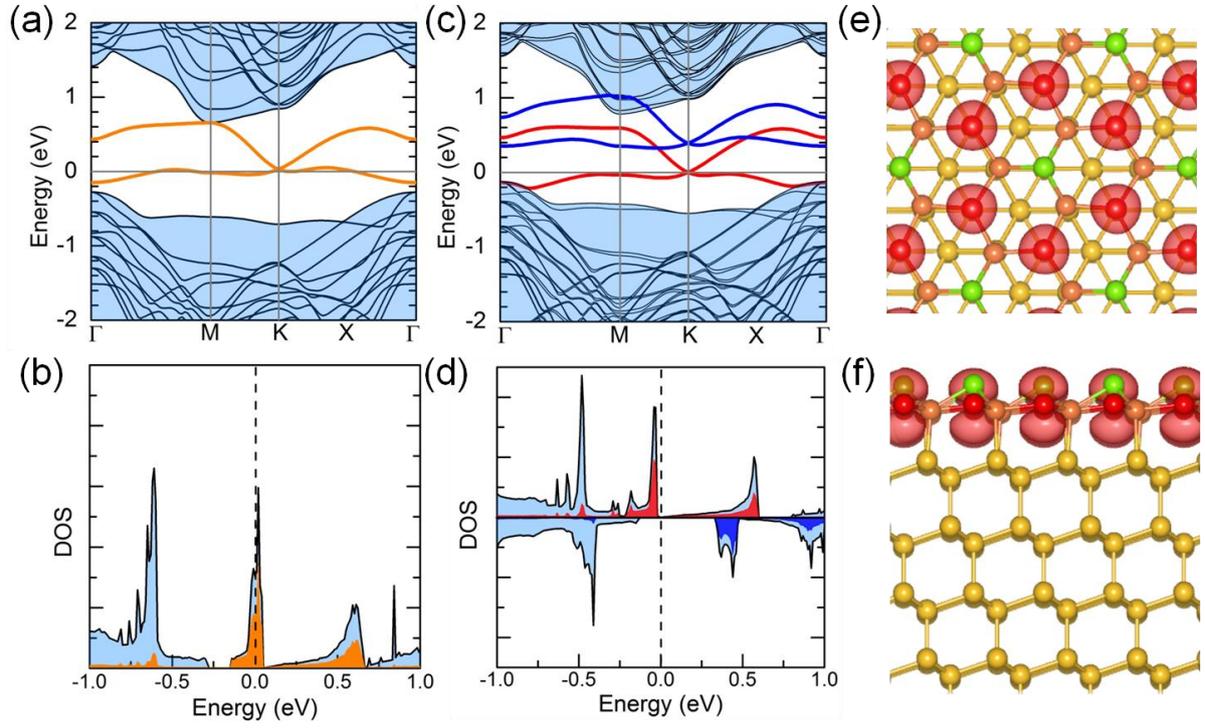

**Figure 2.** Band structures for Si-$\sqrt{3}$ surface in the NM (a) and FM phase (c). The shadow regions are the bulk states and the flat band from Si$_A$ atoms. The yellow, red and blue lines are the Dirac bands from the states of Si$_B$ atoms. (b, d) The corresponding DOS for the Si thin films. The yellow, red and blue regions correspond to the projected DOS on Si$_B$ atoms. (e, f) Top and side view of the spin density distribution at the isosurface level of 0.002 e$^-$/Å$^3$.

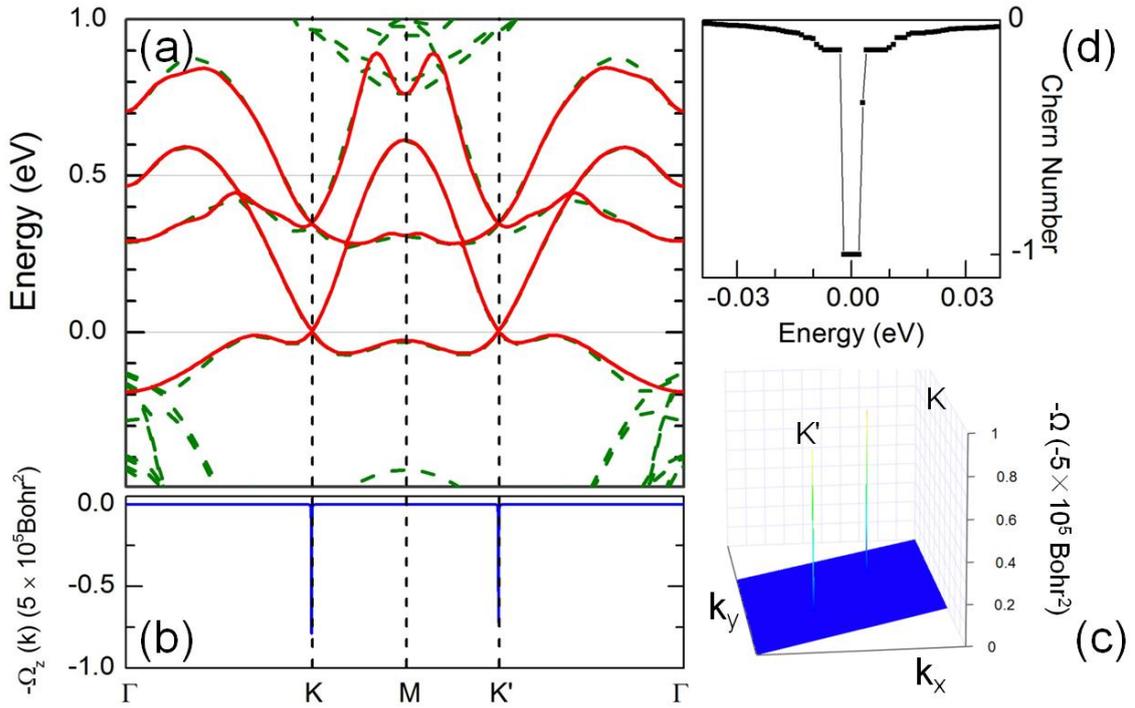

**Figure 3.** (a) Energy band structure for Si-√3 surface calculated by DFT (green dashed line) and fitted by Wannier basis set (red line). (b) The Berry curvature for the valence bands below the Fermi level along the high symmetry direction. (c) The 2D distribution of Berry curvature for the occupied bands in the momentum space. (d) The energy dependence of first Chern number around the quantized conductance plateau.

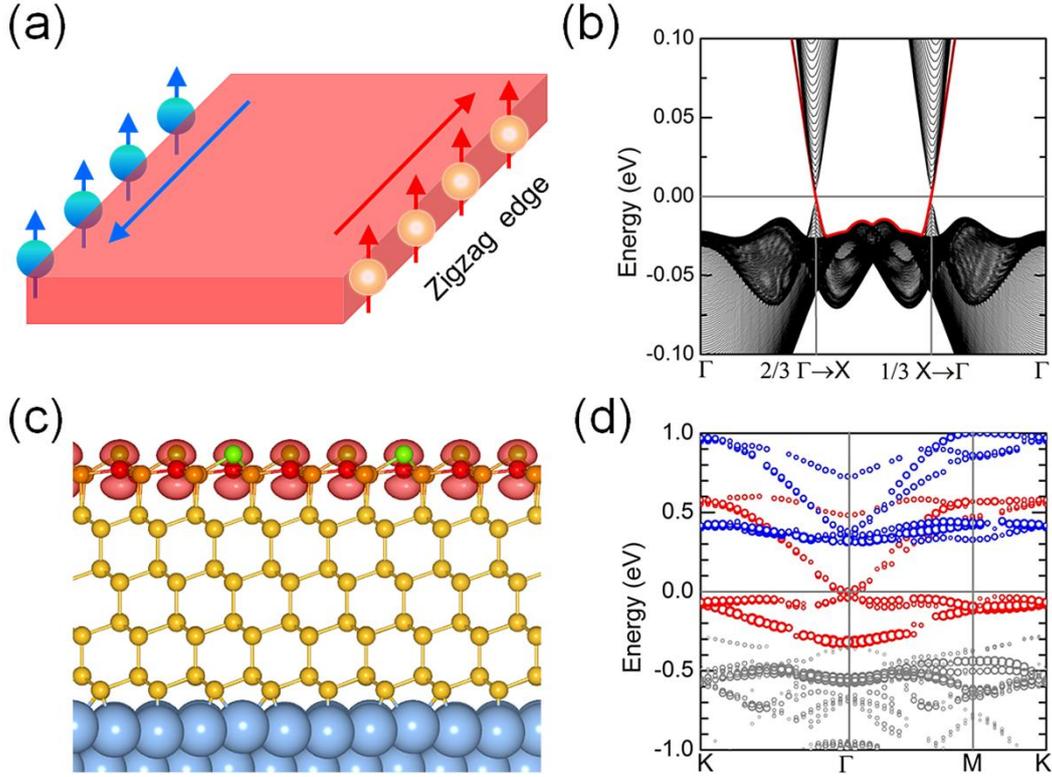

**Figure 4.** (a) Schematic illustration of quantum anomalous Hall conductivity $\sigma_{xy}^H$. The spin-polarized Hall currents along two edges are shown by blue and red color. (b) The tight-binding band structure for Si-$\sqrt{3}$ nanoribbon with the width of 200 zigzag Si chains. The quantized edge states are shown by red lines. (c) Atomic structure and spin distribution of five-layer Si(111) films with $\sqrt{3}$ surface on Ag(111) substrate. The lattice constant of supercell is 11.56 Å. The spin density distribution at the isosurface level of 0.002 e$^-$/Å$^3$ is shown. (d) Projected band structure on the topmost Si atomic layer for the five-layer Si(111) films on silver substrate. The red and blue dots present spin-majority and spin-minority Dirac states from Si$_B$ atoms, respectively. The sizes of dots denote the weight of contribution.